\documentclass[twocolumn,floatfix,showpacs,preprintnumbers,amsmath,amssymb]{revtex4}

\usepackage{graphicx}
\usepackage{dcolumn}
\usepackage{bm}
\usepackage{booktabs}

\begin{document}
\title{Order--disorder induced magnetic structures of FeMnP$_{0.75}$Si$_{0.25}$}

\author{Matthias Hudl}
 \email{matthias.hudl@angstrom.uu.se}
 \affiliation{Department of Engineering Sciences, Uppsala University, Box 534, SE-751 21 Uppsala, Sweden}
\author{Per Nordblad}
 \affiliation{Department of Engineering Sciences, Uppsala University, Box 534, SE-751 21 Uppsala, Sweden}
\author{Torbj\"orn Bj\"orkman}
 \affiliation{Department of Physics and Materials Science, Uppsala University, Box 530, SE-751 21 Uppsala, Sweden}
\author{Olle Eriksson}
 \affiliation{Department of Physics and Materials Science, Uppsala University, Box 530, SE-751 21 Uppsala, Sweden}
\author{Lennart H\"aggstr\"om}
 \affiliation{Department of Physics and Materials Science, Uppsala University, Box 530, SE-751 21 Uppsala, Sweden}
\author{Martin Sahlberg}
 \affiliation{Department of Materials Chemistry, Uppsala University, P.O. Box 538, SE-751 21, Uppsala, Sweden}
\author{Levente Vitos}
 \affiliation{Department of Physics and Materials Science, Uppsala University, Box 530, SE-751 21 Uppsala, Sweden
 and School of Industrial Engineering and Management, Royal Institute of Technology, SE-100 44 Stockholm}
\author{Yvonne Andersson}
 \affiliation{Department of Materials Chemistry, Uppsala University, P.O. Box 538, SE-751 21, Uppsala, Sweden}

\date{\today}

\begin{abstract}
We report on the synthesis and structural characterization of the magnetocaloric FeMnP$_{0.75}$Si$_{0.25}$ compound. Two types of samples (as quenched and annealed) were synthesized and characterized structurally and magnetically. We have found that minute changes in the degree of crystallographic order causes a large change in the magnetic properties. The annealed sample, with higher degree of order is antiferromagnetic with a zero net moment. The as-quenched sample has a net moment of 1.26 $\mu_B$/f.u. and ferrimagnetic-like behavior. Theoretical calculations give rather large values for the Fe and Mn magnetic moments, both when occupied on the tetrahedral and pyramidal lattice site. The largest being the Mn moment for the pyramidal site reaches values as high as 2.8 $\mu_B$/atom.  
\end{abstract}


\maketitle


\section{Introduction}

Compounds based on Fe$_2$P gain increased interest due to a possible application in magnetocaloric refrigeration. Recent publications by Br\"uck et al.\cite{Bruck:1,Bruck:2}, Dagula et al. \cite{Dagula:1} and Cam Thanh et al.  \cite{Thanh:1} showed a huge magnetocaloric effect in FeMnP$_{1-x}$As$_{x}$, FeMnP$_{0.5}$As$_{0.5-x}$Si$_{x}$, and FeMnP$_{1-x}$Si$_{x}$ respectively close to room temperature. The compound FeMnP$_{1-x}$Si$_{x}$ is of particular interest since it consists of non-toxic elements. One drawback of FeMnP$_{1-x}$Si$_{x}$ in regard to applications is a strong thermal hysteresis when undergoing a first order para- to ferromagnetic phase transition. An explanation for the occurrence of a first order transition in Fe$_2$P is given by Yamada and Terao \cite{Yamada:1}. 

\begin{figure}[htbp] 
   \centering
   \includegraphics[width=0.40\textwidth]{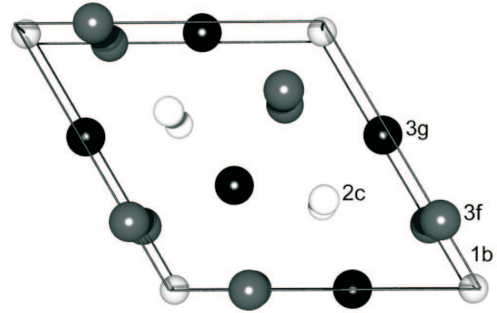}
   \caption{Fe$_{2}$P structure with iron atom positions 3f (dark grey) and 3g (black), and phosphorus atom positions 1b (light grey) and 2c (white).}
   \label{Fe2P:lattice}
\end{figure}

The Fe$_2$P compound has been intensely studied during the last 5 decades. Fe$_2$P crystallizes in a hexagonal structure with space group D$^{3}_{3h}$ (P$\bar{6}$2m) \cite{Rundqvist:1}. The iron atoms occupy two different crystal sites, the 3f-site with four phosphorus atoms surrounding one iron atom (referred to as type-I or tetrahedral site) and 3g-site with 5 phosphorus atoms surrounding one iron atom (referred to as type-II or pyramidal site). The phosphorus atoms occupy the two dissimilar sites 2c (type-I) and 1b (type-II). Each Fe(I) site is surrounded by two P(I) and two P(II) atoms whereas Fe(II) is surrounded by four P(I) and one P(II) atoms (Fig. \ref{Fe2P:lattice}). As regards the magnetic properties Fe$_2$P undergoes a first order para- to ferromagnetic phase transition with a Curie temperature of T$_{C}$ $\approx$ 216 K (see e.g. W\"appling et al. \cite{Wappling:1}, Fujii et al. \cite{Fujii:1}, and Lundgren et al.\cite {Lundgren:1}). It is worth to notice that prior to this investigation structural and magnetic studies on (Fe$_{1-y}$Mn$_{y}$)$_{2}$P and Fe$_{2}$P$_{1-x}$Si$_{x}$ were published by Srivastava et al. \cite{Srivastava:1} and Jernberg et al. \cite{Jernberg:1}. Due to its interesting magnetic properties, Fe$_2$P has also attracted theoretical interest, e.g. as revealed in Refs. \cite{Eriksson:1,Unknown:1}. 

The preliminary phase diagram for FeMnP$_{1-x}$Si$_{x}$ is given by Cam Thanh et al. \cite{Thanh:1} and indicates a structural phase transition from orthorhombic to hexagonal for a silicon content of approximately x=0.25. In spite of prior studies, the importance of the iron to manganese ratio as well as the distribution of those atoms within the Fe$_2$P structure seem rather undefined. The present manuscript is a first attempt towards answering such questions. In our study the magnetic, structural and electronic properties of the FeMnP$_{0.75}$Si$_{0.25}$ alloy have been investigated, using XRD, M\"ossbauer spectroscopy, and magnetic measurements combined with theoretical calculations. We observe a significant difference in the magnetic order depending on the applied sample treatment which is shown to be reversible. The remarkable change in the magnetic order is supposedly caused by the degree of crystallographic order of iron and manganese atoms.


\section{Experimental Details and Methods}

The FeMnP$_{0.75}$Si$_{0.25}$ sample was prepared by a drop synthesis method using a high frequency induction furnace \cite{Rundqvist:2}. The synthesis was done under argon atmosphere and temperatures of approx. 1350 $^{\circ}$C. The first sample was directly taken from the cooled melt and investigated by X-Ray diffraction (XRD). Thereafter some phosphorus was added and the fabricated material was annealed for 10 days at 1000 $^{\circ}$C. All subsequent heat treatments did not involve changes in the element composition. 

The XRD measurements were performed using a focusing Bragg-Brentano type powder diffractometer with CuK$_{\alpha1}$ radiation. The magnetic properties of all samples where investigated by means of DC magnetization measurements mainly using a commercial vibrating sample magnetometer (Quantum Design PPMS). Zero field cooled (ZFC) and field cooled (FC) measurement protocols for different fields were applied. The annealed sample was remelted using an arc melting furnace and quenched. This remelt sample was characterized by magnetic measurements and room temperature M\"ossbauer spectroscopy. The M\"{o}ssbauer spectra were recorded in the absorption mode with constant-acceleration drive and a $^{57}$CoRh source. 

Finally the remelt sample was annealed again and characterized by magnetic measurements and M\"ossbauer spectroscopy. The magnetic properties of the later two samples were the same as those of the originally prepared samples before and after annealing, respectively. 

Calculations using the local density approximation (LDA) have been performed for three different phases of FeMnP. We assumed that all these phases have the hexagonal Fe$_{2}$P structure, with the same crystallographic parameters as observed experimentally for FeMnP$_{0.75}$Si$_{0.25}$. In this structure there are two possible completely ordered phases, one with Mn atoms occupying the pyramidal (high moment) position and one where Mn occupies the tetrahedral (low moment) position. These structures are labeled Mn pyramidal and Fe pyramidal, respectively. In the third phase the Mn and Fe atoms are completely randomly distributed on the two Fe positions, and we will refer to this phase as Disordered.


\section{Results and Discussion}

\subsection{X-ray diffraction}

\begin{figure}[htbp] 
   \centering
   \includegraphics[width=0.45\textwidth]{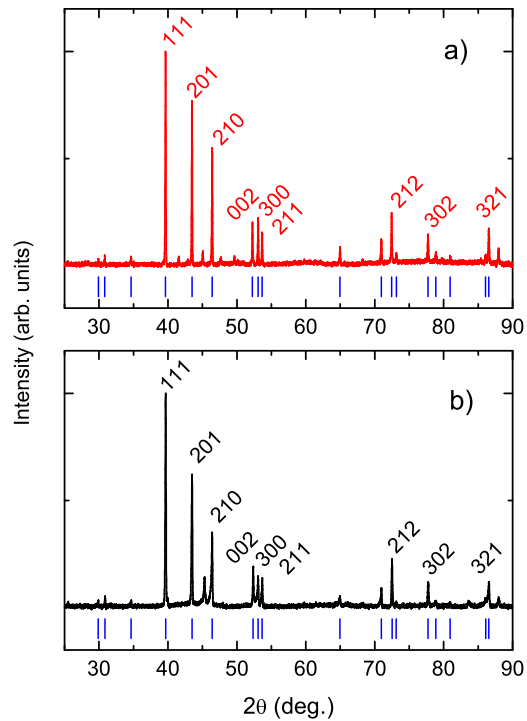}
   \caption{(Color online) XRD patterns of a) an annealed and b) an as-quenched FeMnP$_{0.75}$Si$_{0.25}$ sample. The blue lines mark the indexed peaks of the FeMnP$_{0.75}$Si$_{0.25}$ phase.}
   \label{FeMnPSi:XRD}
\end{figure}

The X-ray data at room temperature confirmed a hexagonal Fe$_{2}$P-type structure with unit cell dimensions a = 5.973 \AA, and c = 3.498 \AA\ for the annealed sample as seen in Fig. \ref{FeMnPSi:XRD} a. The unit cell dimensions for the quenched sample were a = 5.974 \AA, and c = 3.493 \AA. A weak cubic-structure with cell dimension a = 5.653 \AA\ was detected in the raw synthesized material, possibly of Fe$_{3}$Si type. In order to eliminate this fractional phase and to compensate a possible loss of phosphorus during the synthesis some phosphorus was added and the sample was annealed. After the annealing no trace of the Fe$_{3}$Si impurity phase could be deduced from the XRD data. A composition analysis carried out using a electron probe micro-analyzer (WDS-EPMA) as well as energy dispersive spectroscopy (EDS) indicates an excess of iron to manganese with a Fe/Mn ratio of $\sim$ 1.24. In addition our analysis revealed a new impurity phase which amount to $\sim$ 5 \% of our sample consisting of Fe and Si. This impurity phase is also present in the XRD dataset and was determined to be of FeSi type.

\subsection{M\"ossbauer measurements}

\begin{figure}[htbp] 
   \centering
   \includegraphics[width=0.45\textwidth]{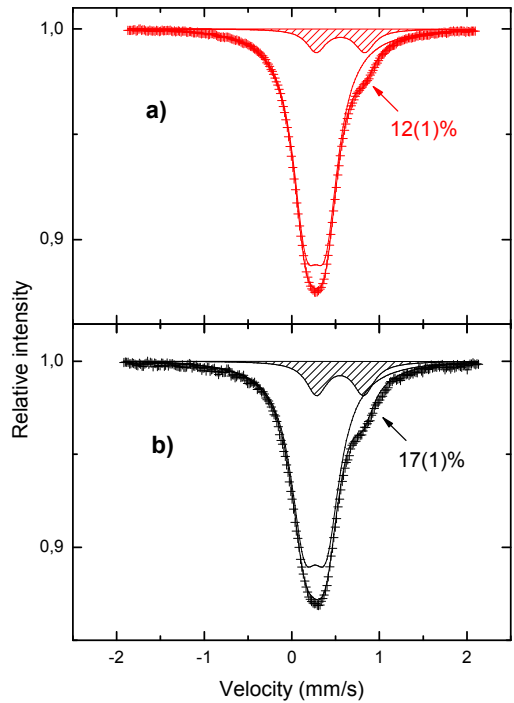}
   \caption{(Color online) Room temperature M\"ossbauer spectra of FeMnP$_{0.75}$Si$_{0.25}$. Spectrum a) corresponds the annealed sample and spectrum b) to the quenched sample. The shaded doublets emanate from the pyramidally coordinated Fe nucleus at the 3g site.}
   \label{FeMnPSi:Moessbauer}
\end{figure}

The quenched sample and the annealed sample were probed by room temperature M\"ossbauer spectroscopy. The distribution of iron atoms on the two inequivalent atomic positions 3f and 3g was investigated. In a Fe$_{2}$P structure the pyramidal 3g site is preferentially occupied by the less electro-negative atom and for small differences in the electronegativity by the atom with larger radius \cite{Furchart:1}. For FeMnP the tetrahedral site is occupied by iron and the pyramidal site by manganese \cite{Srivastava:1}. In the case of FeMnP$_{0.75}$Si$_{0.25}$ with the same stoichiometric number of iron and manganese atom one therefore would expect the iron atoms to be on the tetrahedral site and the manganese atoms on the pyramidal site. The spectral intensities from the M\"ossbauer analysis are modified from the site abundancies due to the so called thickness effect. In the present case the thickness effect for the two spectra can be assumed to be very similar since the total absorbance is almost the same (same M\"ossbauer thickness) and also due to that the lines emanating from site Fe(I) and Fe(II) are not well resolved. The site abundancies of iron atoms on the pyramidal 3g site for the quenched sample is found to 17(1)\%. After annealing the iron concentration on the pyramidal site decreased to 12(1)\% (Fig. \ref{FeMnPSi:Moessbauer}). The separation of Fe and Mn atoms on the two sites accords with the expected preferences.  Already in the quenched sample a significant deviation from random occupany is found and after annealing an added fraction of atoms has diffused to their preferred positions. Any ordering of the pnictide elements on the two phosphorus sites 1b and 2c could not be investigated, in the present study, but may also take place as an effect of the annealing. An elemental P(I)/As substitution has e.g. been found for the closely related compound Fe$_{2}$P$_{1-x}$As$_{x}$ \cite{Catalano:1}. The isomer shifts $\delta$ (mm/s) vs. $\alpha$-Fe and electric quadrupole splittings $\Delta$ (mm/s) do not change significantly due to the metal element ordering, being ($\delta$,$\Delta$)=(0.27(1),0.24(1)) for Fe(I) and (0.55(1),0.53(1)) for Fe(II), respectively.

\subsection{First principles theory}

\begin{figure}[htbp] 
   \centering
   \includegraphics[width=0.45\textwidth]{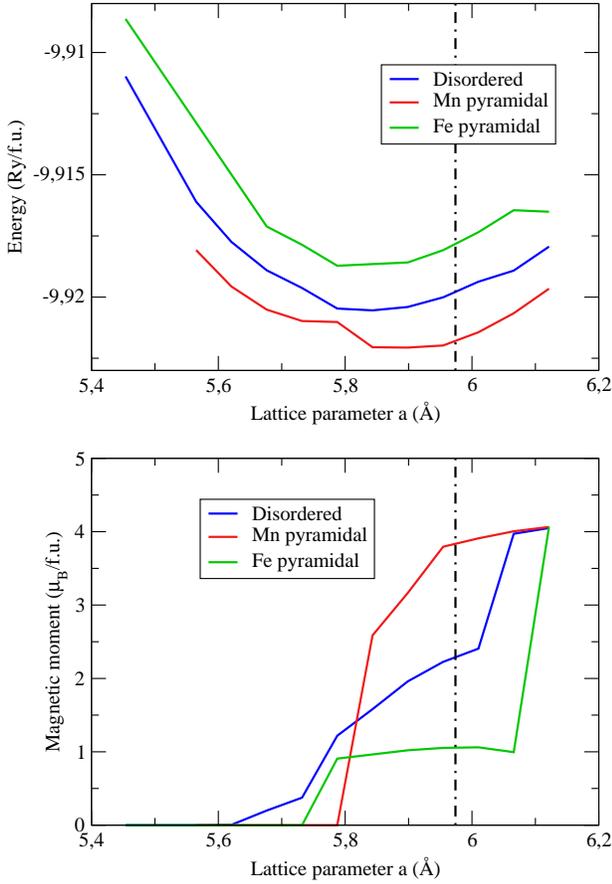}
   \caption{(Color online) Total energy (upper frame) and ordered magnetic moments per formula unit (lower frame) for ordered and disordered phases of hexagonal FeMnP as functions of the lattice parameter. The ordered phases, correspond to the Mn atom occupying the pyramidal (high moment) and tetrahedral (low moment, labeled Fe pyramidal) positions, respectively. In the disordered phase the two positions are randomly occupied by Mn and Fe atoms. The dashed-dotted vertical line indicates the experimental lattice parameter.}
   \label{FeMnP:eos_mom}
\end{figure}

\begin{figure}[htbp] 
   \centering
   \includegraphics[width=0.45\textwidth]{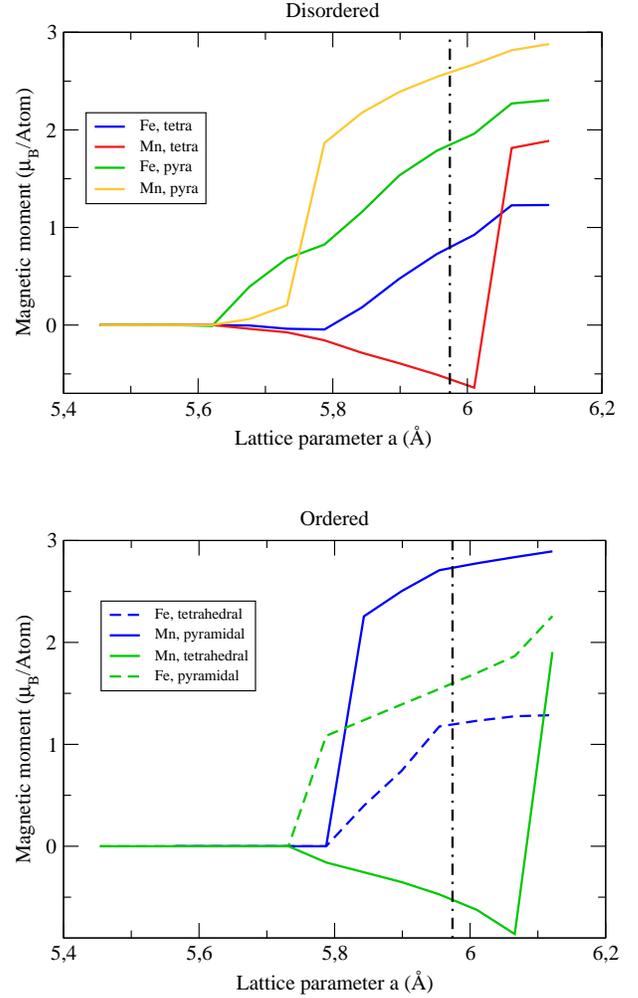}
   \caption{(Color online) The site-projected moments of the ordered and disordered phases of FeMnP as a function of the lattice parameter. The dashed-dotted vertical line indicates the experimental lattice parameter.}
   \label{FeMnP:sitemoms}
\end{figure}

In Fig. \ref{FeMnP:eos_mom} we show the calculated total energy as a function of the lattice parameter, $a$, for three different phases, Mn pyramidal, Fe pyramidal and Disordered (upper panel). As is clear from the figure the Mn pyramidal phase has lowest energy for all considered volumes (or lattice parameters). The energy difference between the different phases is actually rather significant (5-10 mRy/formula unit), and the phase with lowest energy is the one which is also observed experimentally, with Mn preferentially occupying the high moment site. The Mn pyramidal phase has a shallow energy minimum for a lattice constant of 5.83 \AA\textless a\textless5.97\AA, whereas our experimental value is 5.97 \AA\ for the hexagonal FeMnP$_{0.75}$Si$_{0.25}$.

All three curves in the upper panel of Fig. \ref{FeMnP:eos_mom} have a small kink at a lattice constant close to 5.8 \AA. This kink is intimately connected to a sharp change in the calculated total magnetic moment per formula unit, as shown in the lower panel of Fig. \ref{FeMnP:eos_mom}, due to what is known as the magneto-volume effect. For the Mn pyramidal phase, the magnetic moment drops from a considerable value at a lattice constant larger than 5.8 \AA, to vanish completely for lower lattice constants. For the Fe pyramidal phase the transition is not as sharp and the moment does not disappear until a lattice constant of $\sim$5.75 \AA\ is reached, and for the Disordered phase the moment disappears at a lattice constant of $\sim$5.62 \AA. Since the change in the magnetic moment is sharpest for the Mn pyramidal phase, it is natural that the kink in the total energy curve is most pronounced for this phase. The disappearance of the magnetism for lower volumes is a consequence of the competition between kinetic energy of the electron states, which is always lower for a spin-degenerate system, and the exchange energy, which is lower for a spin-polarized system. With decreasing volume the band-width becomes broader, and consequently the kinetic energy becomes the dominating term in the total energy. Hence, lower volumes favor the spin-degenerate state with a vanishing magnetic moment. The competition between kinetic and exchange energy depends intricately on the details of the electronic structure and since the different phases considered in Fig. \ref{FeMnP:eos_mom} have different electronic structures, the transition to a spin-degenerate, non-magnetic state is different for the different phases.

In Fig. \ref{FeMnP:sitemoms} we display the site projected magnetic moments of the Fe and Mn atoms, for the Disordered phase (upper panel) and the Mn pyramidal and Fe pyramidal phases (lower panel). For the Disordered phase we observe that for large volumes all moments are ferromagnetically coupled, but for lower volumes the Mn moment on the low moment site (tetrahedral site) couples antiferromagnetically to the other moments. The transition from ferromagnetic coupling to antiferromagnetic coupling for this atomic moment occurs at a volume very close to the experimental volume. At this volume the Mn-Mn distance on the tetrahedral sites is 2.65 \AA. The distance between the tetrahedral and pyramidal site is similar.

The same behavior is actually exhibited by the Fe pyramidal phase, in that the Mn moment on the tetrahedral site changes from ferromagnetic to antiferromagnetic with decreasing volume (see the lower panel of Fig. \ref{FeMnP:sitemoms}). However, for the Mn pyramidal phase this does not happen, all moments are ferromagnetically coupled. This phase corresponds to Mn atoms occupying the pyramidal, high moment site, with Mn moments approaching 3 $\mu_B$/atom. In the Fe pyramidal phases the Mn moments are always lower than 1 $\mu_B$/atom for a\textless6.1\AA, and it is tempting to explain the stabilization of the Mn pyramidal phase to be due to the exchange energy of the larger Mn moment of the pyramidal site.

\subsection{Magnetization measurements}

\begin{figure}[htbp] 
   \centering
   \includegraphics[width=0.45\textwidth]{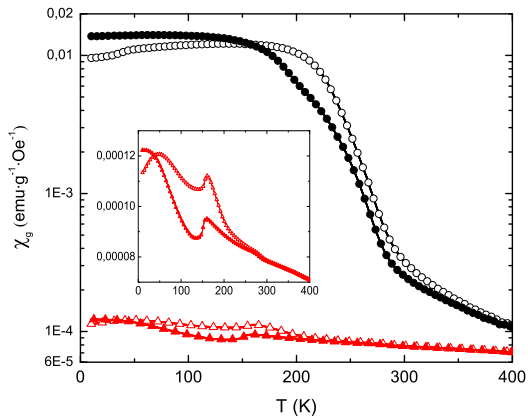}
   \caption{(Color online) Susceptibility (M/H) vs. temperature for FeMnP$_{0.75}$Si$_{0.25}$ measured on the quenched sample in an applied field of 50 Oe (black circles) and on the annealed sample under an applied field of 500 Oe (red triangles, inset). The open symbols indicate measurements using a zero field cold (ZFC) protocol and the filled symbols using a field cooled (FC) protocol.}
   \label{FeMnPSi:MM}
\end{figure}

DC magnetization measurements on the quenched and annealed samples are displayed in Fig. \ref{FeMnPSi:MM}. The as-quenched sample show a broad para- to ferromagnetic phase transition at approx. 250 K accompanied by a strong thermal hysteresis starting already around 280 K. The observed thermal hysteresis is a indicator of a first order nature of the phase transition.

The susceptibility vs. temperature curve for the annealed sample shows a para- to antiferromagnetic phase transition at approx. 160 K. Below the maximum signaling the antiferromagnetic transition at 160 K, the susceptibility slightly increases due to not fully compensated antiferromagnetism. Additionally, there is a significant thermal hysteresis between the ZFC and FC curves; the hysteresis first appears around 280 K, i.e well above the N\'{e}el temperature, but at a temperature that coincides with the temperature for onset of thermal hysteresis on the quenched ferromagnetic sample. It is also of interest to note that this temperature marks the onset of a frequency dependent ac-susceptibility, that remains frequency dependent only down to the antiferromagnetic ordering temperature at 160 K. This indicates that clusters of ferromagnetic order starts to form around 280 K (the same temperature range in which the quenched sample starts to form a long ranged ferromagnetic phase). On further cooling global antiferromagnetic interaction forces the sample in into a long ranged antiferromagnetic order below about 160 K.

It is worth to be mentioned that the closely related compound FeMnP (orthorhombic Co$_{2}$P structure) also show an antiferromagnetic structure for 176K\textless T\textless265K with a doubling of the crystallographic c axis. Below 175 K a complicated modulated helical antiferromagnetic structure is developed \cite{Sjostrom:1}. FeMnP is a fully ordered compound with tetrahedral Me(I) site and the pyramidal Me(II) site fully occupied by Fe and Mn atoms, respectively. 

\begin{figure}[htbp] 
   \centering
   \includegraphics[width=0.45\textwidth]{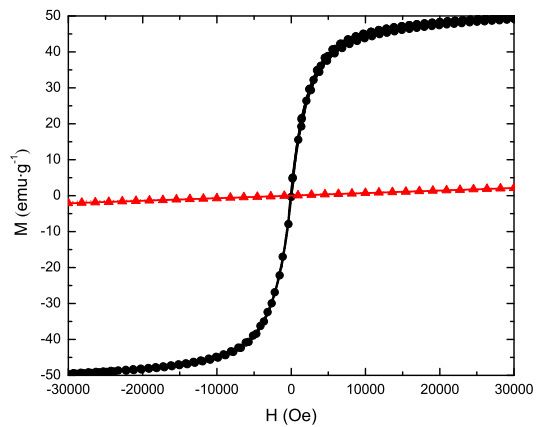}
   \caption{(Color online) Magnetization vs. applied field for an as-quenched sample at 30 K (black circles) and an annealed sample at 30 K (red triangles).}
   \label{FeMnPSi:MM-Hyst}
\end{figure}

The magnetic moment of both samples as a function of the applied field at 30 K is shown in Fig. \ref{FeMnPSi:MM-Hyst}. At an applied field of 3 T and 30 K the measured magnetic moment is 1.26 $\mu_{B}$ per f.u. for the quenched sample and 0.05 $\mu_{B}$ per f.u. for the annealed sample. The figure distinctly pictures the transformation of the low temperature state of the material from ferromagnetic to antiferromagnetic by only an almost marginal change of the site occupancy of the Fe and Mn atoms. 

\begin{figure}[htbp] 
   \centering
   \includegraphics[width=0.45\textwidth]{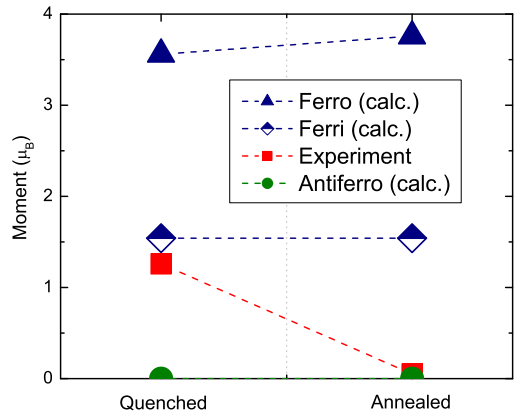}
   \caption{(Color online) Experimental magnetic moments/f.u. in comparison with calculated moments assuming ferro--, ferri-- and antiferromagnetic ordering.}
   \label{FeMnPSi:Exp-Theo}
\end{figure}

In Fig. \ref{FeMnPSi:Exp-Theo} a comparison between calculated and measured magnetic moments is shown. The calculated magnetic moments are obtained by averaging the calculated ordered moments for the Mn pyramidal, Fe pyramidal, and Disordered phase in proportion to the Fe/Mn ratio and measured disorder from M\"ossbauer spectroscopy. For the ferrimagnetic case an Fe/Mn ratio of 1 and full crystallographic order was assumed. It can be seen that the experimental results for the as-quenched sample and the annealed sample do not coincide with ferromagnetic ordering. Both samples, annealed and as quenched exhibit a more complex magnetic structure such as antiferromagnetism or ferrimagnetism. 

In the case of Fe$_{1-x}$S a rather similar experimental result was explained by the ordering of vacancies on magnetic sublattices, see Takayama et al.\cite{Takayama:1}. The appearance of an antiferromagnetic phase in Fe$_{2-x}$P due to heat treatment as seen for FeMnP$_{0.75}$Si$_{0.25}$ was not observed,  see e.g. Lundgren et al. \cite{Lundgren:1}. The influence of vacancies due to non-full site occupation was therefore not further studied.


\section{Summary and Conclusions}

In this manuscript we report on the synthesis and structural characterization of FeMnP$_{0.75}$Si$_{0.25}$, a compound which crystallizes in the hexagonal Fe$_{2}$P-type structure. Two types of samples (as quenched and annealed) were synthesized and characterized structurally (using M\"ossbauer spectroscopy) and magnetically. It is found that marginal changes in the degree of crystallographic order causes a large change in the magnetic properties. Annealing causes a larger degree of order compared to the rapidly quenched sample, and our analysis from the M\"ossbauer data suggests that for the annealed sample $\sim$ 12 \% of the pyramidal (high moment) site is occupied by Fe atoms and $\sim$ 88 \% by Mn atoms. This correspond to almost full order for the measured Fe/Mn ratio of $\sim$ 1.24. Assuming full site occupation this means that the tetrahedral site has $\sim$ 88 \% Fe atoms and $\sim$ 12 \% Mn atoms.  For the quenched sample the corresponding numbers are $\sim$ 17 \% of the pyramidal site is occupied by Fe atoms and $\sim$ 83 \% by Mn atoms. These small changes, which eventually may be accompanied by a pnictide element ordering, result in drastically different magnetic behaviour. 

The annealed sample, with higher degree of order is antiferromagnetic with a zero net moment. The as-quenched sample has a net moment of 1.26 $\mu_B$/f.u. Our experimental finding that the magnetism depends very delicately upon the degree of order is in qualitative agreement with the theoretical first principles results. 

The theoretical calculations give rather large magnetic moments for the Fe and Mn atoms, both when occupied on the tetrahedral and pyramidal site. The largest being the Mn moment for the pyramidal site, which reaches values as high as 2.8 $\mu_B$/atom. We do not have atomic resolved experimental values to compare these theoretical values with, but we can note that normally theory and experiment agree with each other for atomic projected moments of magnetic materials (see e.g. Ref.\cite{Eriksson:1}) with an error being less than 10 \%. If we assume that this is also the case for the currently studied system, we must conclude that the as-quenched sample is a ferrimagnet or a non-collinear magnetic structure, possibly involving a spin-spiral state, since a ferromagnetic coupling of the calculated atomic moments would result in a net moment of $\sim 3.5$ $\mu_B$/f.u., a value much larger than the measured value (see Fig. \ref{FeMnPSi:Exp-Theo}). 

The observed magnetic response of FeMnP$_{0.75}$Si$_{0.25}$ is strongly dependent on the proportion of Fe and Mn atoms occupying the tetrahedral and pyramidal sites. A very small increase of the Fe concentration on the pyramidal site, only a few percent, causes a major change from an antiferromagnet to a ferrimagnet with a rather large saturation moment. It is unclear if a modification of the stoichiometry would cause a similar change in the magnetic response, but it is tempting to speculate that this may be the case. We also find from our theory that an ordered phase with all Mn atoms on the pyramidal site and all Fe atoms on the tetrahedral site has a significantly lower energy compared to the disordered phase. 

The here studied material, FeMnPSi, has been characterized structurally and magnetically, with a range of experimental techniques and by first principles theory.  As member of a family of materials which are relevant for magnetocaloric refrigeration. Our study indicates that the influence of crystallographic order and disorder on the magnetocaloric properties is important and should be studied in more detail. This involves both varying the concentration of Fe and Mn as well as different annealing conditions. It is also desirable to undertake neutron scattering experiments to detect atom projected moments. Such studies are underway.


\section{Acknowledgments}

Financial support from the Swedish Energy Agency (STEM) and the Swedish Research Council (VR) is acknowledged.
Calculations done on supercomputer resources provided by SNAC. We are grateful to Ece G\"ul\c{s}en for the sample fabrication, 
Rebecca Bejhed, Hans Harrysson, and Hans Annersten for composition analysis, and Roland Mathieu for valuable suggestions and advices.


\bibliographystyle{apsper}

\end{document}